\documentclass{article}
\usepackage[utf8]{inputenc}
\usepackage{float}
\usepackage{amsmath}
\usepackage{setspace}
\usepackage{upgreek}
\usepackage{siunitx} 
\usepackage{graphicx}
\usepackage{amsmath, amssymb}
\usepackage{authblk}
\usepackage{url}
\usepackage[margin=3cm]{ geometry }
\title{ Differential Sensing Approaches for Scattering-Based Holographic Encryption }
\date{}
\author[1] { Mohammadrasoul Taghavi}
\author[1] { Edwin A. Marengo}
\affil[1]  { Department of Electrical and Computer Engineering, Northeastern University, Boston, Massachusetts 02115, USA}

\begin{document}

\maketitle

\noindent{\bf {\large Abstract}}: We develop a new scattering-based framework for the holographic encryption of analog and digital signals. 
The proposed methodology, termed ``differential sensing'', involves encryption of a wavefield image by means of two hard-to-guess, complex and random scattering media, namely, a background and a total (background plus scatterer) medium. Unlike prior developments in this area, not one but two scattering media are adopted for scrambling of the probing wavefields (as encoded, e.g., in a suitable ciphertext hologram) and, consequently, this method offers enhanced security. In addition, 
while prior works have addressed methods based on physical imaging in the encryption phase followed by computational imaging in the decryption stage, we examine the complementary modality wherein encryption is done computationally while decryption is done analogically, i.e., via the materialization of the required physical imaging system comprising the ciphertext hologram and the two unique (background and total) media. The practical feasibility of the proposed differential sensing approach is examined with the help of computer simulations incorporating multiple scattering.  The advantages of this method relative to the conventional single-medium approach are discussed for both analog and digital signals. 
The paper also develops algorithms for the required {\em in situ} holography as well as a new wavefield-nulling-based approach for scattering-based encryption with envisioned applications in real-time customer validation and secure communication. 

\section{Introduction}
The encryption of sensitive information—including passwords, barcodes, images, signatures, fingerprints, and other biometric data—along with the real-time validation and identification of individuals, products, packages, and documents like passports and credit cards, is of critical importance in banking, homeland security, and numerous other fields. Holography has been traditionally of substantial relevance in this area \cite{glass2012holographic,su2012security,javidi2000securing,park2011applications,haleem2022holography,lancaster1995future,ferri2002biometric, javidi2021roadmap,chen2022compact,wang2022differentiable}. Since the introduction of double random phase encoding (DRPE) \cite{refregier1995optical,unnikrishnan2000optical,chen2014advances,tajahuerce2000optoelectronic}, optical cryptography has garnered significant interest thanks to its unprecedented capability for handling high-dimensional and parallel operations \cite{liu2010double,chen2010optical}. As DRPE has evolved, it has been adapted to various mathematical frameworks, such as the Fresnel transform, Gyrator transform, and fractional Fourier transform, each offering tailored benefits for different encryption needs \cite{situ2004double,matoba1999encrypted,singh2009gyrator}. Beyond these, optical encryption strategies that rely on imaging technologies—such as ghost imaging, diffractive imaging, and interferometric imaging—have demonstrated their potential in safeguarding information \cite{clemente2010optical,cao2022physically}. However, despite these innovations, DRPE and its variations are not impervious to security threats \cite{carnicer2005vulnerability}. Studies have highlighted their susceptibility to several attack types, including chosen-ciphertext, chosen-plaintext, and ciphertext-only attacks \cite{peng2006known,peng2006chosen,liu2015vulnerability}. These attack methods have been continually refined and are crucial in evaluating the robustness of optical encryption systems. Additionally, concerns linger regarding the security of computer-generated holography (CGH)-based optical encryption systems, as their vulnerabilities have yet to be fully explored, raising questions about their viability in real-world applications \cite{zhou2020learning,zhou2020vulnerability}. Given the vulnerabilities and limitations of secure holography based on two-dimensional phase screens, as well as the practical challenges associated with constructing DRPE systems, researchers have shifted their focus toward alternative approaches, being of enormous interest in recent years the exploitation of speckle-inducing multiple scattering phenomena. 

Since the hologram's image depends on the illumination field, the latter can play the role of encryption key while the hologram acts as ciphertext or {\em vice versa}. In recent years large key spaces have become attainable via the exploitation of as many wave degrees of freedom as possible, including orbital angular momentum (OAM) (see, e.g., \cite{yang2023angular,zhang2022holographic,ren2020complex, cao2024accurate}), thereby rendering increased capacity and security. Moreover, the field propagating into the hologram can itself be encrypted in a number of ways, e.g., by superimposing a space light modulator (SLM) (which governs the incident field) and a programmable metasurface (which unscrambles the field so as to generate the sought-after field at the hologram) \cite{qu2020reprogrammable}. The latter methodology can be further improved by incorporating polarization control \cite{yu2022dynamic,yu2023ultrahigh}. Another approach that has received a lot of attention is the inclusion of complex, multiple scattering media in the generation of ciphertexts. Work in the latter direction constitutes a pivotal reference point for the developments of the present paper. The idea is that a complex scattering medium can be adopted to scramble the information-carrying waveform in complicated, multipath ways that are hard to decode without knowledge of the medium or of decryption (image-formation) models or ``keys'' based on it. 
This approach has been demonstrated in a number of studies including 
Liu et al. \cite{liu2020exploiting} where only part of the scattering medium is used to generate the ciphertext while decryption is done computationally via the speckle-correlation scattering matrix
(SSM) method. 
Recently, deep learning has revolutionized the field of optical imaging by offering advanced solutions for complex tasks like image reconstruction and denoising, outperforming traditional methods. By learning from vast amounts of data, deep learning models can recover hidden targets through scattering media, though they often require large and comprehensive datasets to achieve robust and generalized performance.
The work of Zhao et al. \cite{zhao2022speckle} adopts deep learning in the decryption phase,  and demonstrates application of scattering-based encryption to 
face recognition. The paper by
Zhou et al. \cite{zhou2020learning} also develops a learning model as the decryption key. 
The holographic encryption method in Yu et al. \cite{yu2022scattering} employs both a scattering medium and chaos-based mathematical encryption. It requires the physical presence of the complex medium in both encryption and decryption, where the latter step relies on re-focusing via phase conjugation.  

In this paper we build upon the pioneering work in these \cite{liu2020exploiting,zhao2022speckle,zhou2020learning,yu2022scattering} and other related papers \cite{qinghan_new,chen2016optical,xiao2022physically} and propose a number of system and algorithmic additions, through which the practical applicability of scattering-based encryption systems can be further enhanced. While prior developments consider the adoption of a single scattering medium, in the present work we examine the possibility of implementing the encryption via two different scattering media. The associated approach, which we term ``differential sensing'', enhances the performance characteristics. In particular, it further increases the key space and renders more immunity to certain eavesdropping attacks. 
Moreover, while the focus in some of the prior work has been on analog encryption, via the physical medium, followed by digital decryption via deep learning or other methods, our focus is on the complementary approach where the encryption step can be done either analogically or digitally (in the latter case, via the singular value decomposition (SVD)-based inverse filter of the corresponding forward map) while the associated decryption is done in a real-time analog manner, in the presence of the correct scattering media. 

\begin{figure}[H]
\centering
\includegraphics[width=15 cm]{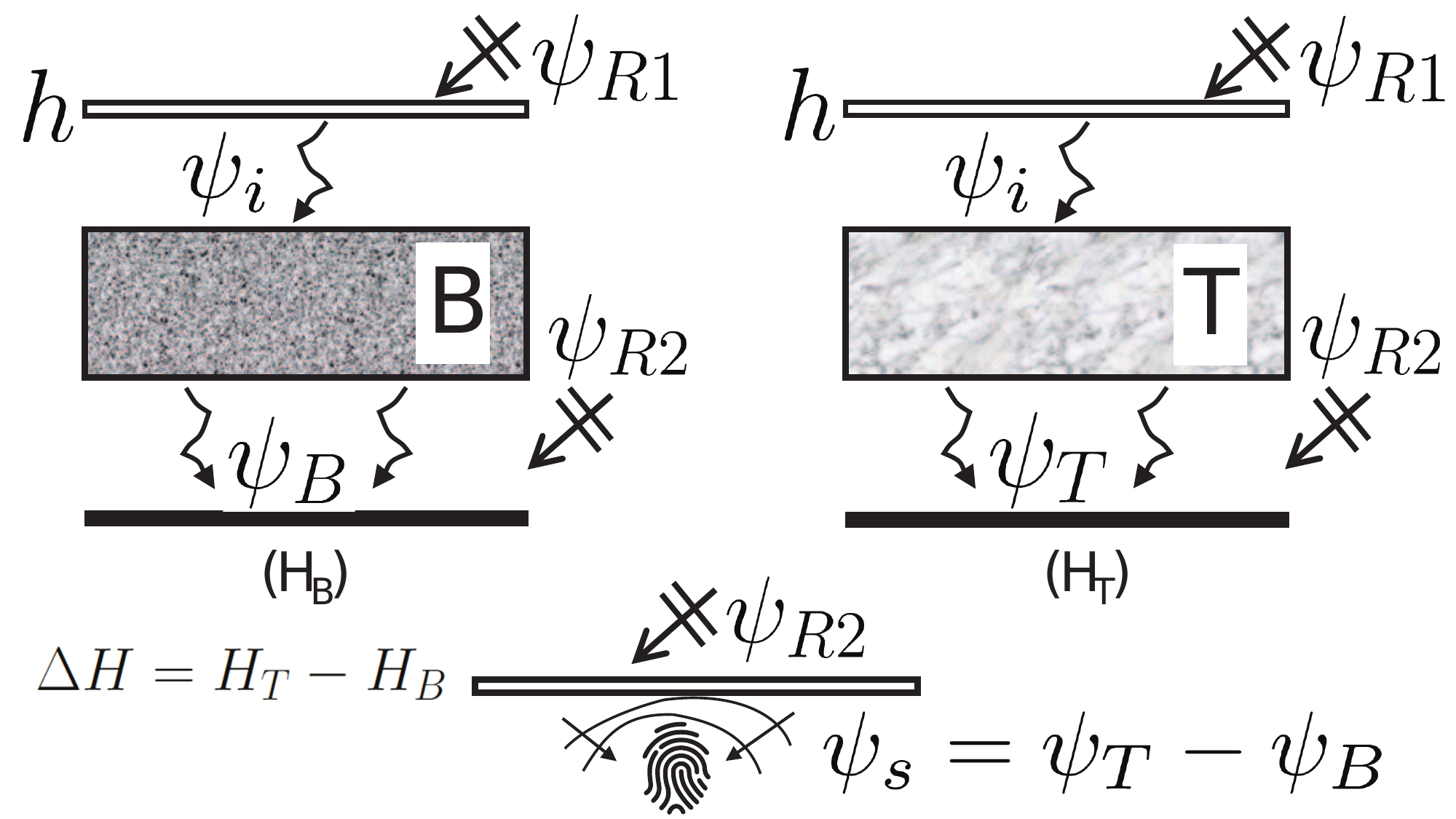}
\caption{Scattering in complex media B and T upon illumination of the incident field hologram $h$ with reference wave $\psi_{R1}$. The figure illustrates the holographic measurement of the resulting scattered fields with the help of another reference wave $\psi_{R2}$, through which the background and total medium response holograms ($H_B$ and $H_T$, respectively) and the corresponding differential sensing hologram ($\Delta H$) are obtained.}
\label{newfigholog1backup}
\end{figure}
Figure~\ref{newfigholog1backup} conceptually illustrates the system under consideration. It consists of 1) a hologram ($h$) (the ciphertext) which carries the information for the synthesis (under illumination wave $\psi_{R1}$) of a hard-to-infer probing wave ($\psi_i$); 2) complex scattering media or ``micromedia'' (labelled B and T in the figure) (the cryptographic key); and 3) the resulting waves $\psi_B$ and $\psi_T$ in media $B$ and $T$, respectively, as measured in a region of interest (ROI) where sensors (including sensors with holographic, phase-reading capabilities) can be placed. The corresponding output holograms ($H_B$ and $H_T$) and the corresponding differential sensing hologram ($\Delta H=H_T-H_B$) facilitate under the correct illumination (say $\psi_{R2}$) the reproduction of the sought-after information-carrying fields. In the differential sensing approach, the relevant information (e.g., a fingerprint, as illustrated in the figure) is carried by the corresponding scattered field $\psi_s$. The idea is that sole access (e.g., by a potential hacker or eavesdropper) to the ciphertext hologram carries little or no information in the absence of the required complex scattering media (B and T) since they are essential in the physical materialization of the required image-formation channel. Both media are required for successful signal retrieval. Thus the latter play, jointly, the role of cryptographic key. Incidentally, the illumination waveforms $\psi_{R1}$ and $\psi_{R2}$ may too be highly specialized for the sought-after reconstruction, thereby providing yet another avenue for added channel security. 

Importantly, the envisaged scattering-based holographic encryption has a multitude of functionalities, through which both analog and digital signals can be securely stored. In the secure storage channel application the signal is a field image that cannot be retrieved without the correct media. Only the intended receiver who has the correct scattering media can recover the correct signal. Another application is customer validation that must be performed once (single-attempt) in real time (at checkpoint) in front of a suitable authority or validating system. In this context both required media may remain under the customer's control. Alternatively, for enhanced security, the customer may have access to one of the micromedia while the validating authority controls access to the other one. In this case immunity to eavesdropping is ensured so long as access to only a single medium precludes the pertinent image reconstruction. In this application, the relevant analog signal may be, for example, a person's fingerprint, or another unique identifying feature, e.g., a signature or passcode. For further complexity, the hidden signal can be a further processed version of the ``field image'' resulting from a complex set of non-adaptive and hard-to-guess queries pertinent to the relevant identifying feature. This provides an additional degree of security since in this framework an eavesdropper would need to have simultaneous access, in real time, to the hologram, the medium, the unique identifying feature, as well as a passcode needed to reproduce the queries or challenges giving rise to the correct signal. 
Another application, which is pertinent for both validation and secure communication, is the use of the medium itself as a digital signal, e.g., a bar code. 
In this case the hidden information is encoded in the medium, as a digital sequence of scattering pixels or voxels (``1'' if occupied by the scattering center and ``0'' if empty). A number of options emerge. For example, the output signal may be known to be a null or low-amplitude image due to interference or nulling effects that arise when the correct medium is placed in the neighborhood of the imaging plane. If such null effects are extremely unlikely with the incorrect media, then this provides a method for user validation or as a concept for the selective ingestion of valid bar codes (representing, e.g., data or cryptographic keys) out of large sets or blocks. We develop in this work general methods and algorithms pertinent to all of these envisioned applications, contrasting whenever possible single-medium-based (non-differential-sensing) techniques versus double-media-based (differential-sensing) methods. 

We wish to highlight that the technologies for the physical realization of the envisioned complex micromedia are already available. For instance, remarkable progress has been made in the field of metamaterials, particularly in manipulating electromagnetic waves at subwavelength scales \cite{engheta2006metamaterials}. 
Metamaterials have found diverse applications in the field of secure holography, broadly categorized into two types. The first involves the use of two-dimensional metamaterials, known as metasurfaces, which can act as holograms themselves \cite{genevet2017recent,zheng2015metasurface,liang2024single,yuchong2023multifunctional,li2023metasurface,zheng2021metasurface,liu2024single,liu2022phase}. These metasurfaces can be designed to be either static or spatially or temporally reconfigurable, allowing for dynamic control in CGH \cite{li2017electromagnetic,barati2020time,barati2020topological}. This approach leverages the unique properties of metasurfaces to manipulate light at the nanoscale, enabling the creation of highly sophisticated and customizable holographic images. The second application involves the use of three-dimensional volumetric metamaterials, such as clusters of nanoparticles distributed within a volume, which serve as scattering media for encryption and imaging purposes \cite{kadic20193d,da2024asymmetric,tanzid2018absorption}. These scattering media can create complex scattering environments that significantly enhance the security of holographic encryption by making it challenging to decipher the encoded information without precise knowledge of the medium's configuration \cite{popoff2010image, popoff2010measuring}.

The remainder of the paper is organized as follows. Section 2 presents the formulation of the proposed differential sensing method along with a discussion of alternative forms of the required {\em in situ} (in the medium) holographic measurements. The proposed methodology is illustrated in section 3 with computer simulations incorporating the critical multiple scattering effects. We comparatively discuss results pertinent to the single-medium and differential sensing architectures for both analog and digital signals.
We also discuss a new field-nulling method for checkpoint validation and communication which can be implemented in both single-medium and differential sensing systems. This technique leads, naturally, to single-pixel imaging architectures that are relevant for applications where the information is retrieved dynamically, via varying sequences of the hologram and micromedia. Section 4 provides concluding remarks. 

\section{Methods and Algorithms}

We consider in the following the use of two complex, random media for the encryption of holographic information. One of these media can be regarded as the reference or background medium, for the purposes of the associated scattering formulation. We shall refer to this first medium as ``background'' or medium B. The second medium can be thought of as being composed of the background plus a medium perturbation or ``scatterer''. We shall refer to the latter as the ``total medium'' or medium T. It is assumed that these scattering materials are confined within a scattering region  $\tau$. Next we discuss propagation and scattering of wavefields in these media within the scalar wave framework, and introduce the singular value decomposition (SVD) of the map from the source induced in the hologram (in the hologram region $\Sigma$) to the resulting field in the sensing region $\Omega$ where sensors are placed.

Let $G_B({\bf r},{\bf r'})$ be the background Green's function governing radiation in medium B, where ${\bf r}$ and ${\bf r'}$ are position vectors in the relevant configuration space. When the probing hologram in $\Sigma$ is illuminated by reference wave $\psi_{1R}$ (see Fig.~\ref{newfigholog1backup}) an incident field ($\psi_i$) is generated which is used to probe the medium. Within Rayleigh-Sommerfeld diffraction, this field can be described as arising from a surface source $\rho_h$ in $\Sigma$ that is proportional to the boundary value of the incident field $\psi_i$ at $\Sigma$; in particular, under unity obliquity factor 
\begin{equation}
    \rho_h({\bf r})=2ik \psi_i({\bf r}) \quad {\bf r}\in \Sigma.\label{eq_aug_8_2024_1}\end{equation}
This enables us to formulate the map from the hologram signal to the generated field in the medium as a map from an equivalent surface source $\rho_h({\bf r})$ in the hologram surface $\Sigma$ to the resulting field $\psi_B({\bf r})$ at the observation or sensing region $\Omega$. Within this framework we define the linear map $P_B$ from the hologram source to the resulting field in the sensing region  as
\begin{equation}
\psi_B({\bf r}) =(P_B \rho_h)({\bf  r}) =  I_{\Omega}({\bf r}) \int_{\Sigma} d{\bf r'} G_B({\bf r},{\bf r'}) \rho_h({\bf r'}) \label{eq_june27_2024_1}\end{equation}
where $d{\bf r'}$ denotes spatial differential element and $I_{B}({\bf r})$ is an indicator or masking function whose value is equal to one if ${\bf r}\in \Omega$ and is zero otherwise. We assume next that the field measurements are taken with an array of $N_D$ detectors, and in this case the support $\Omega$ consists of the set of sensor positions, say ${\bf R}_j,j=1,2,\cdots,N_D$. Moreover, we also assume that the hologram is formed by a number ($N_h$) of pixels, associated to positions ${\bf X}_i,i=1,2,\cdots,N_h$, and within this discrete data context the map $P_B$ in (\ref{eq_june27_2024_1}) takes the form of a response matrix which we call $K_B$. The latter characterizes the map from the vector of hologram pixel sources (to be denoted as $\bar \rho_h$) to the vector of sensor field data (to be denoted as $\bar\psi_B$). We conveniently adopt this framework next and introduce the corresponding SVD. In particular, 
\begin{eqnarray}
    \bar\psi_B &=& K_B \bar\rho_h \nonumber \\
    &=&\sum_{n=1}^{N_B} \sigma_{n}^{(B)} \bar\psi_{n}^{(B)} \bar\rho_n^{(B)\dagger} \bar\rho_{h} \label{eq_june27_2024_3}\end{eqnarray}
  where $\bar\rho_n$ and $\bar\psi_n$ are the right-hand and left-hand singular vectors, respectively, while $\sigma_n$ are the singular values, $N_B={\rm min}(N_D,N_h)$, and $\dagger$ denotes the conjugate transpose. As is well known, the inversion of this map, to synthesize a source ($\bar\rho_h'$) which generates a given field ($\bar\psi_B'$), may be achieved in practice via the regularized pseudoinverse, i.e.,  
    \begin{equation}
        \bar\rho_h' = \sum_{\sigma_{n}^{(B)}\geq\varepsilon } \bar\rho_n^{(B)} \bar\psi_n^{(B)\dagger}
\bar\psi_B' /[\sigma_n^{(B)}]^2 \label{july8_2024_2}
    \end{equation}
where $\varepsilon>0$ is a regularization parameter that is chosen in a tradeoff between solution accuracy and stability. 

The same formulation applies, of course, to the total medium, where the background Green function $G_B$ is substituted by the total medium's Green function $G_T$. Thus in analogy to (\ref{eq_june27_2024_1}) the map $P_T$ from the hologram source to the total field $\psi_T$ in the sensing region is given by 
\begin{equation}
\psi_T({\bf r}) =(P_T \rho_h)({\bf  r}) =  I_{\Omega}({\bf r}) \int_{\Sigma} d{\bf r'} G_T({\bf r},{\bf r'}) \rho_h({\bf r'}).
\label{eq_aug_8_2024_3}
\end{equation}
Similarly, the medium $T$'s counterparts of the discrete formulation expressions (\ref{eq_june27_2024_3}) and (\ref{july8_2024_2}) are
\begin{eqnarray}
    \bar\psi_B &=& K_T \bar\rho_h \nonumber \\
    &=&\sum_{n=1}^{N_T} \sigma_{n}^{(T)} \bar\psi_{n}^{(T)} \bar\rho_n^{(T)\dagger} \bar\rho_{h} \label{eq_june27_2024_3b}\end{eqnarray}
    and
    \begin{equation}
        \bar\rho_h' = \sum_{\sigma_{n}^{(T)}\geq\varepsilon } \bar\rho_n^{(T)} \bar\psi_n^{(T)\dagger}
\bar\psi_T' /[\sigma_n^{(T)}]^2. \label{july8_2024_2partb}
    \end{equation}

For a given scattering medium in $\tau$, the field at the sensing region $\Omega$ is the sum of the incident field (in free space) coming from the holographic source at $\Sigma$ plus the scattered field (in free space) coming from the secondary source that is induced in said medium. Importantly, the incident field component is always present, in the associated total field, regardless of the particular scattering medium. This is an unwanted characteristic, since it can give rise to information leakage if the information-carrying hologram is hacked or intercepted. In that case, an eavesdropper could retrieve at least the portion of the total field associated to the hologram-induced probing field. A more resilient approach is to encode the information only in the scattered field, since then access to the hologram alone is insufficient for the retrieval of the scattered field signal which depends in a complex way on both the probing field and the unknown complex medium. Within this methodology one could, of course, encode the information signal or image in the scattered field, relative to the free space background. This simple method has, on the other hand, the drawback that if the medium were hacked or intercepted, then the eavesdropper would automatically know the critical systemic key of the entire holographic system, namely, the scattering response matrix, through which it could be possible to gain at least partial information (e.g., features such as the relevant signal subspace, etc.) about the hidden signal. We propose a yet more robust methodology, wherein said scattering is relative not to free space but to another complex medium which functions as the background. In this approach access to the hologram or to one of the materials alone does not reveal the stored signal since in this case the input waveform (probing field in the background) is itself scrambled by the background while the systemic constituent or key is the scatterer or ``perturbation'' relative to the background which is itself unspecified unless one has access to both media, an unlikely scenario. 

In particular, this ``differential sensing'' approach is based on the scattering response matrix, which is given by 
\begin{equation}
K_s=K_T-K_B .\label{eq_aug_8_2024_2}\end{equation}
It governs the map from $\bar\rho_h$ to the scattered field in the sensing region, which we represent via the vector $\bar\psi_s$. Thus 
\begin{equation}
    \bar\psi_s = K_s \bar\rho_h . \label{eq_aug_8_2024_5}\end{equation}
The corresponding SVD is of the form 
\begin{equation}
    K_s = \sum_{n=1}^{N_s} \sigma_n^{(s)} \bar\psi_n^{(s)} \bar \rho_n^{(s)\dagger}\bar\rho_h , \label{eq_aug8_2024_7}\end{equation}
and the relevant pseudoinverse, for the purposes of designing a source $\bar\rho_h'$ that generates a desired scattered field vector $\bar\psi_s'$, is given by 
\begin{equation}
        \bar\rho_h' = \sum_{\sigma_{n}^{(s)}\geq\varepsilon}  \bar\rho_n^{(s)}\bar\psi_n^{(s)\dagger}
\bar\psi_s' /[\sigma_n^{(s)}]^2. \label{july8_2024_4}
    \end{equation}
 In this framework the space of realizable signals is defined by ${\cal S}={\rm span}(\psi_n^{(s)}; \sigma_n^{(s)}\geq\varepsilon)$, and thus the inversion in (\ref{july8_2024_4}) works well if the sought-after signal $\bar\psi_s'$ does not have significant energy in the corresponding null space ${\cal N}$, where ${\cal S} \cup {\cal N }= \mathbb{C}^{N_D}$. 
 
 In summary, in the proposed differential sensing method media B and T are randomly generated. Their response matrices are measured or computed and the relevant singular system $(\bar\rho_n^{(s)},\bar\psi_n^{(s)},\sigma_n^{(s)})$ and associated SVD for the corresponding scattering response  is determined (eq.(\ref{eq_aug8_2024_7})) . The required holographic source is computed via standard linear inversion (eq.(\ref{july8_2024_4})). The accuracy of the corresponding inversion can be checked and, if needed, the output signal can be redesigned so as to lie within the pertinent space of realizable signals. 
Finally, the relevant CGH is obtained as follows. The incident field at the hologram surface $\Sigma$ (vector $\bar\psi_i'$) associated to source $\bar\rho_h'$ is given from (\ref{eq_aug_8_2024_1}) by 
\begin{equation}
    \bar\psi_i' = \bar\rho_h'/(2ik) .\label{eq_aug_8_2024_14}
\end{equation}
The discrete form of the required hologram transparency $t({\bf r})$ is then defined by 
     \begin{equation}
          t ({\bf X}_i) = | \psi_{R1}({\bf X}_i)+ \bar\psi_i'|^2 \quad i=1,2,\cdots, N_h . \label{eq_aug_8_2024_15}
     \end{equation}
Upon excitation by reference wave $\psi_{R1}$, this transparency generates four field components, one of which is the real image field corresponding to the desired incident field (corresponding to $\bar\psi_i'$), as needed. Through suitable design (e.g., Leith-Upatnieks holography) the impact of the other field components can be kept reduced in the vicinity of the relevant scattering materials (B and T), as desired. 

\subsection{Holographic Measurements}
To implement these methods in the optical regime, where only field intensities are measurable, one needs to consider practical ways to extract the phase information carried by the scattered signals or other (derivative) information-carrying signatures of interest. We conclude this section with a discussion of four approaches to extract information-carrying signals contained in the scattered field. In the first method (labelled ``a'') the background field itself is adopted as reference wave. Thus in the absence of an additional reference wave the resulting holograms associated to the background and total fields, $H_{B}^{(a)}$ and $H_{T}^{(a)}$, respectively, are (apart from a multiplicative constant)
\begin{eqnarray}
H_B^{(a)}&=&|\psi_B|^2 \nonumber \\
H_T^{(a)}&=&|\psi_B|^2+ 2\Re(\psi_B^*\psi_s)+|\psi_s|^2 \label{eq_may11_2024_1}\end{eqnarray}
so that the difference hologram 
\begin{equation}
\Delta H^{(a)}=H_T^{(a)}-H_B^{(a)}=2\Re(\psi_B^*\psi_s)+|\psi_s|^2.\label{eq_may11_2024_2}\end{equation}
If $|\psi_s|<<|\psi_B|$, which holds if the scatterer is a weak perturbation (in the background) then one obtains from eqs.(\ref{eq_may11_2024_1},\ref{eq_may11_2024_2}) the normalized difference hologram
\begin{equation}
\delta H^{(a)} = \frac{\Delta H^{(a)}}{2\sqrt{H_B^{(a)}}} = \Re (\hat\psi_B^* \psi_s) \label{eq_may22_2024_3}\end{equation}
where we have introduced the normalized backbround field $\hat\psi_B=\psi_B/|\psi_B|$. The quantity $\delta H^{(a)}$ can be used as the information-carrying signal ($f_1^{(a)}=\delta H^{(a)}$). In particular, in the signal design stage one selects media B and T and designs for a desired $\psi_s$. Since $\psi_B$ is also preset in this process, one can evaluate $\delta H^{(a)}$ and adopt this quantity as the encrypted output or validation signal. In addition, upon generation (in the presence of the background B) of background field $\psi_B$ it is possible to excite the further normalized hologram $\delta H^{(a)'}=\delta H^{(a)}/\sqrt{H_B}$ to generate the virtual image field 
\begin{equation}f_2^{(a)}=\psi_s+\psi_s^*\left(\psi_B/|\psi_B|\right)^2,\label{eq_may11_2024_6}\end{equation}
 which reveals, apart from a nuisance real-image-part component, the scattered field $\psi_s$. 

In the second method, a reference plane wave $\psi_{R2}$ is used in the hologram-development process. In this case the corresponding holograms are 
\begin{eqnarray}
H_B^{(b)}&=& |\psi_B|^2+2\Re(\psi_{R2}^*\psi_B)+|\psi_{R2}|^2  \nonumber \\ 
H_T^{(b)}&=&|\psi_B+\psi_s|^2+2\Re[\psi_{R2}^*(\psi_B+\psi_s)]+|\psi_{R2}|^2 .\label{eq_may11_2024_4}
\end{eqnarray} 
In practice, the amplitude of $\psi_{R2}$ can be chosen such that $|\psi_B|<<|\psi_{R2}|$ and $|\psi_s|<<|\psi_{R2}|$, and it is easy to show from (\ref{eq_may11_2024_4}) that if $|\psi_s|<<|\psi_B|$ then the corresponding difference hologram is given by 
\begin{equation}
\Delta H^{(b)}=H_T^{(b)}-H_B^{(b)}\simeq 2\Re(\psi_{R2}^*\psi_s).\label{eq_may11_2024_5}\end{equation}
If the hologram is subsequently illuminated with reference wave $\psi_{R2}$ then a virtual-image field emerges, thereby revealing the scattered field $\psi_s$, as desired. As is well known (\cite{goodman2005introduction}, ch.~9), by carefully selecting the direction of propagation of the reference wave, it is possible to isolate the associated virtual and real images, such that a clear image of the scattered field is revealed. 

We discuss next a third possibility, which can be used to obtain multiple images analogous in form to the one in eq.(\ref{eq_may11_2024_6}). To achieve this, we design the reference wave $\psi_{R2}$ such that the scatterer (the perturbation to the background) does not scatter, in the background, upon illumination by such wave. This can be achieved via a general field nulling methodology, as we elaborate later in the paper. In this approach, the corresponding holograms for the background and the total fields are
\begin{eqnarray}
H_B^{(c)}&=&|\psi_{B2}+\psi_B|^2\nonumber \\
H_T^{(c)}&\!\!\!=\!\!\!&\!\!|\psi_{B2}+\psi_B+\psi_s|^2\!\!=\!\!|\psi_{B2}+\psi_B|^2\!+\!2\Re[(\psi_{B2}+\psi_B)^*\psi_s]\!+\!|\psi_s|^2 \label{eq_may11_2024_7}\end{eqnarray}
where $\psi_{B2}$ represents the sum of the reference wave ($\psi_{R2}$) plus the field that is scattered in the background medium upon excitation by said reference wave. It follows from (\ref{eq_may11_2024_7}) that if the scattered wave $\psi_s$ is weak relative to the other fields then 
 the associated difference hologram 
\begin{equation}
\Delta H^{(c)}=H_T^{(c)}-H_B^{(c)}\simeq 2 \Re [(\psi_{B2}+\psi_{B})^*\psi_s].\label{eq_may11_2024_9}\end{equation}
By analogy with the discussion in eq.(\ref{eq_may11_2024_6}) one can then generate the normalized hologram 
\begin{equation}\delta H^{(c)'}=\frac{\Delta H^{(c)}}{2H_B^{(c)}}.\label{eq_may11_2024_10}\end{equation}
 Now, upon generation of the field $\psi_{B2}+\psi_B$ via illumination of the background with the incident field due to hologram $h$ and the reference wave $\psi_{R2}$, and placing hologram $\delta H^{(c)'}$ in the measurement plane, the field 
\begin{equation}
f_3=\psi_s+\psi_s^*(\psi_{B2}+\psi_B)^2/|\psi_{B2}+\psi_B|^2 \label{eq_may11_2024_12}\end{equation}
is launched which carries, along with a nuisance real-image component (the second term in eq.(\ref{eq_may11_2024_12})), the sought-after scattered field.

Furthermore, we can combine the holograms in the first and third methods to obtain yet another image. Unlike the aforementioned methods where weak scattering with respect to the background medium was assumed in order to simplify the results, in the following method this is not required. Thus the following approach, which is differential in nature, holds for more arbitrary scattered signals. In this (fourth) approach, we compute from the measured $\Delta H^{(a)}$ and $\Delta H^{(c)}$ the following differential quantity: 
\begin{equation}
\Delta H^{(c,a)}={\Delta H^{(c)}-\Delta H^{(a)}}= 2 \Re(\psi_{B2}^*\psi_s).\label{eq_may11_2024_15}\end{equation}
Within this approach it is convenient to also measure the intensity of $\psi_{B2}$ by illuminating the background only with the reference wave $\psi_{R2}$. We measure the corresponding intensity at the sensing plane:
\begin{equation}
H_{B2}=|\psi_{B2}|^2.\label{eq_may11_2024_18}\end{equation}
Finally, we evaluate from eqs.(\ref{eq_may11_2024_15},\ref{eq_may11_2024_18}) the normalized hologram 
\begin{equation}
\delta H^{(c,a)}=\frac{\Delta H^{(c,a)}}{2H_{B2}}.\label{eq_may11_2024_21}\end{equation}
Now, if this computed hologram is subsequently placed in the sensing plane, and the background is illuminated with the reference wave $\psi_{R2}$, then the hologram becomes illuminated by the total reference wave $\psi_{B2}$ and therefore it launches the field 
\begin{equation}
f_4=\psi_s+\psi_s^*\left(\psi_{B2}/|\psi_{B2}|\right)^2 \label{eq_may11_2024_23}\end{equation}
which (like $f_2$ and $f_3$) showcases the sought-after scattered field $\psi_s$. It also becomes possible, via averaging of the wavefields $f_2,f_3,f_4$, to enhance the visibility of the sought-after scattered field $\psi_s$ thanks to partial destructive interference of the real-image contributions in $f_2, f_3, f_4$. Thus, despite the presence in these holographic reconstructions of the associated twin images, the latter are expected to take different forms in the different holograms, thereby enabling identification of the information-carrying signal $\psi_s$. It is also important to emphasize that the methodology giving rise to $f_4$ is general, and does not require weak scattering (with respect to the background).  Importantly, the same approach leading to the signature $f_4$ can be implemented for an arbitrarily large number of views (where $\psi_{R2}\rightarrow \psi_{Rn}$, where $n>2$, and the associated $f_4\rightarrow f_{n+2}$), rendering, potentially, a large number of images ($f_1,f_2,\cdots,f_{n+2}$) all of which carry the common information-carrying component of $\psi_s$, as desired. 
We discuss next computer simulations of the differential sensing methods under the assumption that the relevant scattered fields in complex media can be extracted in practice from intensity-only field data via these and other approaches. 
\begin{figure}[H]
\centering
\includegraphics[width=15 cm]{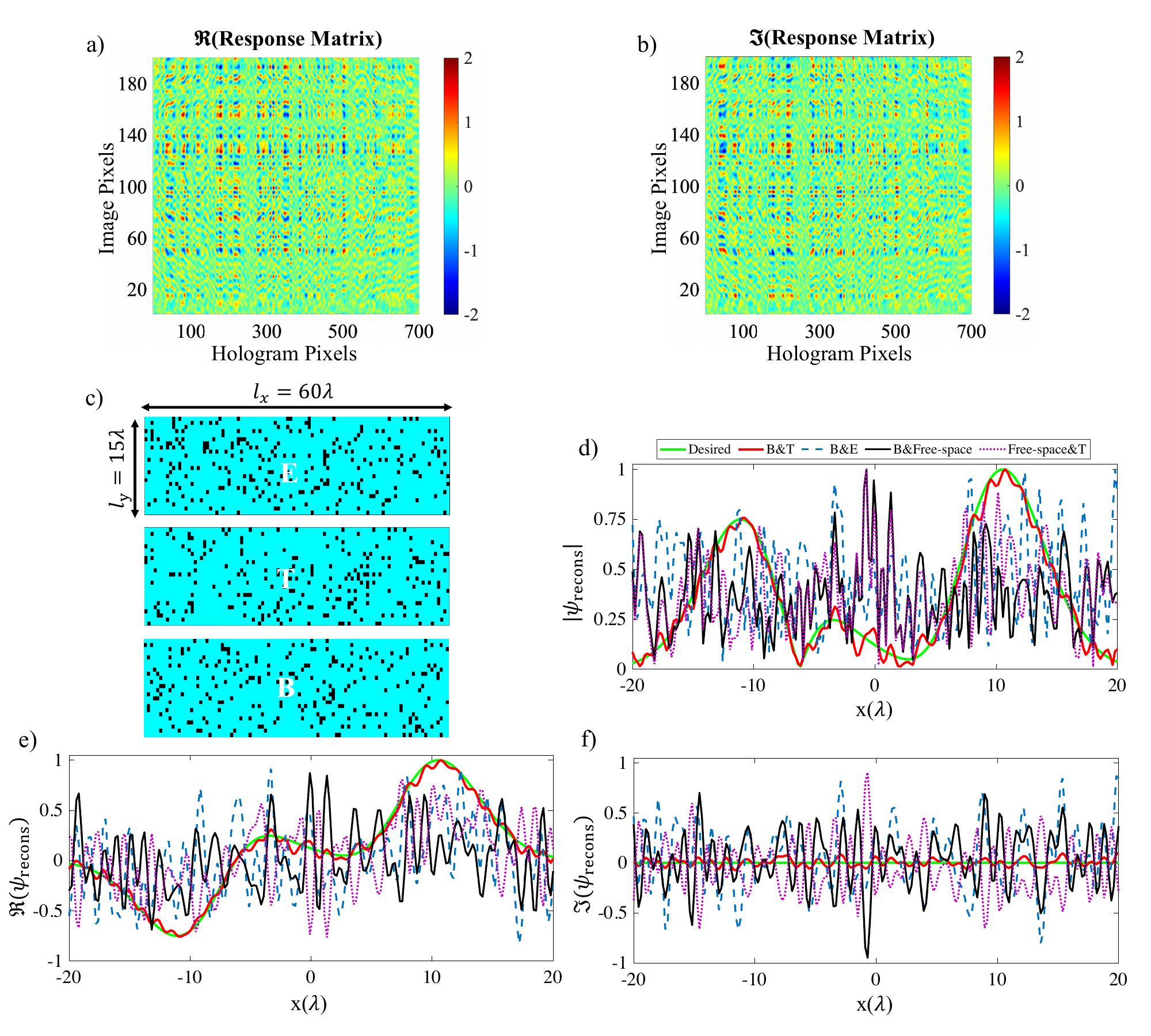}
\caption{Differential sensing method for holographic encryption. (a) Real part of the response matrix for the correct media (materials B and T in part c). (b) Corresponding imaginary part of the response matrix. (c) Background (B), total (T), and eavesdropping (E) media. (d) The amplitude, (e) real, and (f) imaginary parts of the reconstructed signals in different scenarios: Desired (green line), correct background (B) and total medium (T) (red line), correct background medium (B) and incorrect eavesdropping medium (E) (dashed blue line), correct background medium (B) and free space as second medium (black line), and free space as background medium and correct total medium (T) (dashed-purple line).}
\label{fig_4}
\end{figure}
\section{Results and Discussion}

In this section we illustrate the proposed differential sensing methods using computational simulations in two-dimensional (2D) space. The hologram is contained in a circular arc of radius $R_h=200 \lambda$ and angular extent $\theta_h=120^\circ$, while the ROI is a line of length $l_R=40 \lambda$ centered at the origin. The scattering medium is placed at a distance of $10\lambda$ from the ROI. In the first round of simulations we show the effectiveness of the multiple scattering approach in the encryption of arbitrary analog signatures. The results clearly illustrate advantages and disadvantages of the differential sensing approach, in relation to the conventional single-medium approach, and the tradeoffs therein. As a second round of results, we discuss the application of the differential sensing method in the encryption of digital signals such as bar codes. Exhaustive simulation results shed light on the associated encryption performance. Finally, in this section we also illustrate alternative encryption methods based on wavefield nulling. 
\begin{figure}[H]
\centering
\includegraphics[width=15 cm]{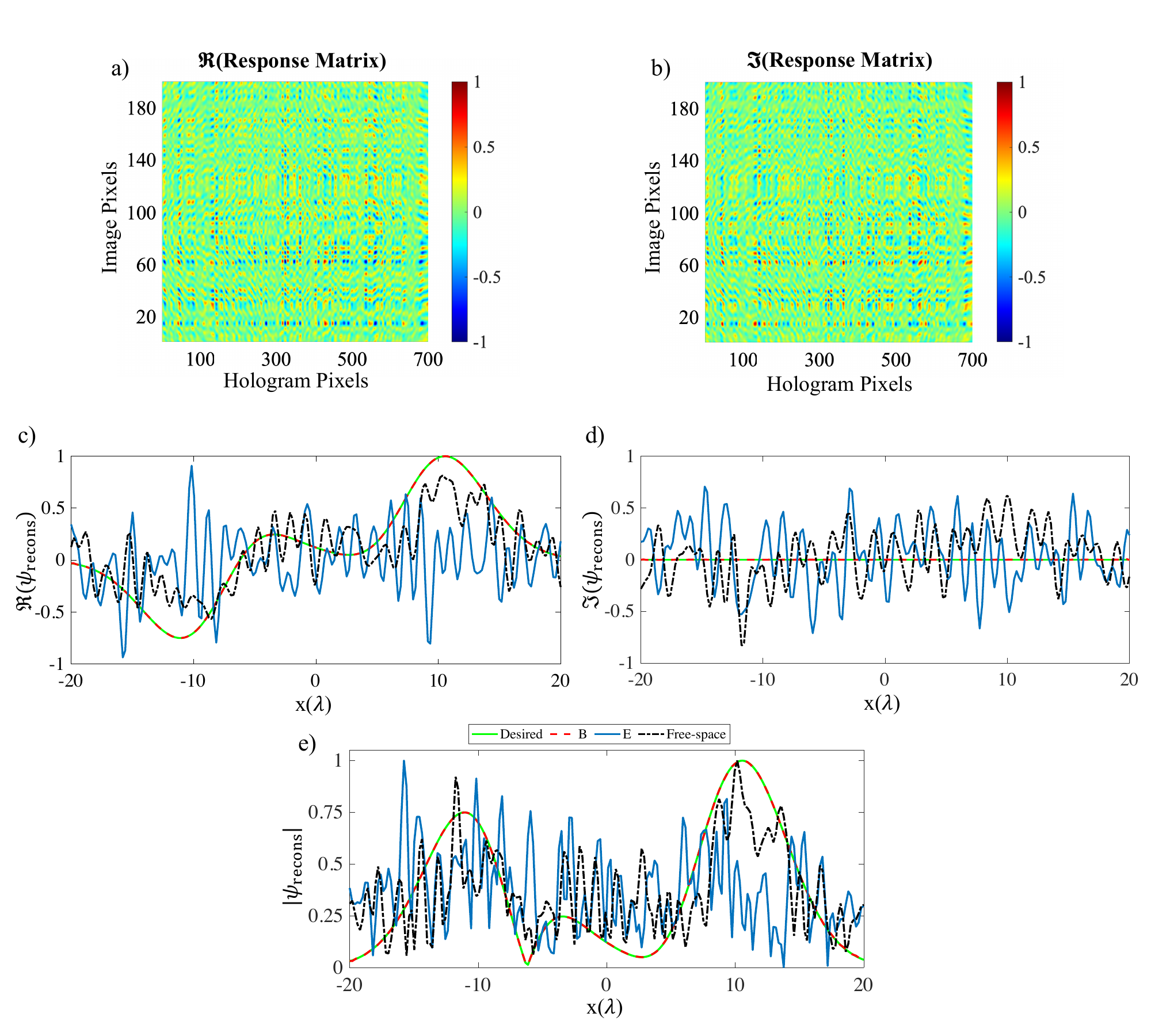}
\caption{Single-medium method for holographic encryption. Parts (a) and (b) show the real and imaginary parts of the generated response matrix using the background medium (B). The (c) real, (d) imaginary, and (e) absolute values of the (normalized) reconstructed signal corresponding to the following scenarios: Desired (green line), correct medium (B) (red line), eavesdropping medium (E) (blue line), and free space (no scattering medium) (black-dashed line). }
\label{fig_5}
\end{figure} 
\subsection{Analog Implementation}
To simulate the required complex multiple scattering media, we considered random aggregates of small scatterers with the same scattering strength (which we chose to be $\tau = -i$ to represent a thin metallic cylinder, see, e.g., \cite{balanis2012advanced}, p.606) whose scattering response was modeled via the Foldy-Lax multiple scattering method~\cite{ishimaru2017electromagnetic,huang2013efficient}. The media 
used in the following simulations consist of $n_x=100$ and $n_y=25$ pixels, with corresponding sizes of $l_x=60\lambda$ and $l_y=15\lambda$. For the sought-after encryption, this corresponds to a key size $K_{s,1}=2^{2500}$ for the conventional single-material methods, and of an expanded key size $K_{s,2}=2^{5000}$ for the proposed differential sensing approach. The density of the population (density of scatterer-occupied pixels) can be controlled during the key generation process and, as discussed further later, plays an important part in the developments. 
In the following simulations, we highlight results for the three random media depicted in Fig.~\ref{fig_4}(c), but we also discuss exhaustive simulations for a large sample of representative media. The media labeled as B and T correspond to the background and total media, respectively, while the medium labeled as E acts as eavesdropper's medium. The points in the material correspond to the aforementioned small scatterers or inhomogeneities. It is assumed, for simplicity, that the index of refraction of the surrounding medium where the scatterers are embedded is equal to one. 

\begin{figure}[H]
\centering
\includegraphics[width=15 cm]{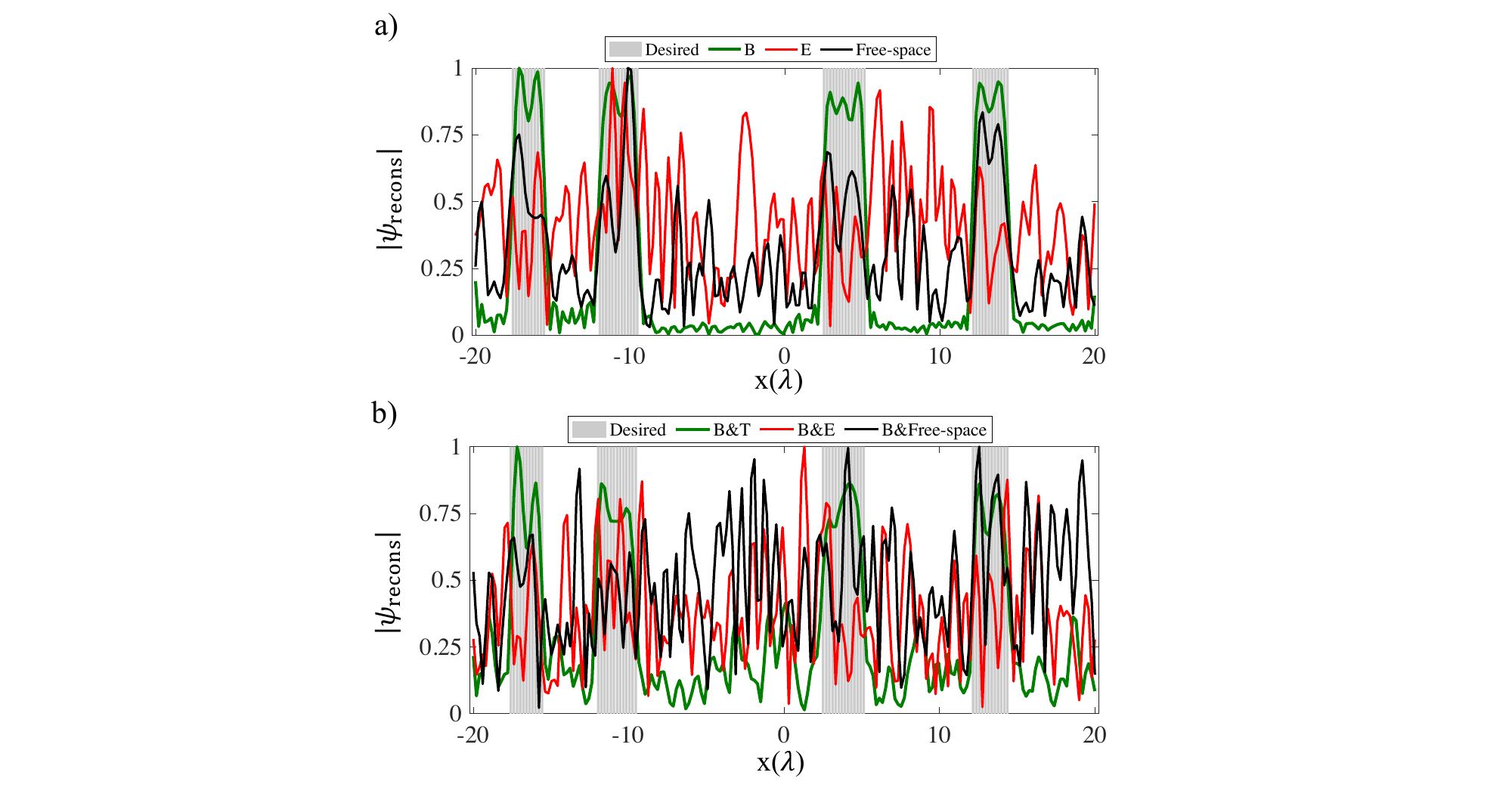}
\caption{(a) Normalized amplitude of the digital signal reconstructions using the single-medium method, where the desired barcode is illustrated with gray bars. Results associated to the following cases are shown: correct medium (B) (green line), eavesdropping medium (E) (red line), and free space (no scattering medium) (black line). (b) Normalized amplitude of the digital signal reconstructions using the difference-based sensing method: Desired barcode (gray bars), correct background (B) and total medium (T) (green line), correct background medium (B) and eavesdropping medium (E) (red line), correct background medium (B) and free-space as second medium (black line). }
\label{fig_6}
\end{figure}

A crucial step in the design process is the generation of the response matrix for the correct media. This can be achieved by calculating the generated fields in the ROI due to multiple hologram excitations or probing fields. This is done separately for both the correct background and total media. The resulting data is used to generate the corresponding forward scattering map, and its highly unique singular system, from which it becomes possible to determine (via classical filtered inversion) a hologram transparency signal giving rise to the sought-after scattered waveform at the ROI. Importantly, signal selection (at the ROI) can be tailored to the unique SVD of the scattering response in order to enhance reconstruction accuracy and resolution. In this simulation we chose the well-known Peaks function in MATLAB as the signal to be reconstructed in the presence of the correct media~\cite{MATLAB}. 
Figures~\ref{fig_4}(a) and \ref{fig_4}(b) show the real and imaginary parts of the response matrix for the media B and T in Fig.~\ref{fig_4}(c). Parts (c), (d), and (e) of Fig.~\ref{fig_4} illustrate the accuracy of the reconstructed scattered signal at the ROI when the correct media B and T are used. These results are representative of many simulations run during the course of this investigation, where consistently good approximations of the desired signals were achieved only in the presence, in the scattering region, of the correct media. To highlight the retrieval of the correct signal shape, in these and other plots in the paper (see, e.g., Fig.~\ref{fig_5} and Fig.~\ref{fig_6}) the corresponding signals are normalized relative to the respective peak values. These figures also illustrate various interception scenarios. 
Under the proposed encryption method, the interceptor does not have access to either the background medium or the total medium. But, as Figs.~\ref{fig_4}(c), \ref{fig_4}(d), and \ref{fig_4}(e)  clearly reveal, poor reconstructions are obtained even in the unlikely scenario in which the eavesdropper gains access to one of the two materials. Reconstructions based on at least one incorrect medium do not reveal the features of the hidden signal and are clearly inaccurate, as desired.

To fully understand the benefits of using a difference-based technique for holographic encryption, it is essential to compare its characteristics with those of the conventional single-medium method. Fig.~\ref{fig_5} illustrates the corresponding single-medium performance for the same media used in Fig.~\ref{fig_4}. In particular, we assumed that the correct medium is medium B. We studied the quality of the reconstructions when medium E and free space act as eavesdropping media. Figures~\ref{fig_5}(a) and \ref{fig_5}(b) show the respective (B medium) response matrix. Parts (c), (d), and (e) of the same figure display the absolute value, real part, and imaginary part, respectively, of the reconstructed images corresponding to different media. 
The red, blue, and black lines in parts (c), (d), and (e) of Fig.~\ref{fig_5} correspond to scenarios where the scattering medium is the correct key (B), the eavesdropper key (E), and free space, respectively. 
The images for the correct medium accurately match the correct output signal.
The most important outcome of this study can be understood by comparing the reconstructed signals in Fig.~\ref{fig_5}, when the utilized medium is free space, to the similar scenarios in Fig.~\ref{fig_4}, where one of the media is free space. Unlike the differential method, the intruder is able to obtain considerable amount of information if no scattering medium is used. This is a serious problem that could limit the utilization of the single-medium method, particularly for simple scattering materials or media of limited spatial dimensions, as we elaborate further next. 

As these results show, one of the most vulnerable scenarios in the single medium method is when the interceptor does not use any scattering material as key, i.e., when free space is used as the eavesdropping medium. It is not surprising that the leakage of signal features tends to be prevalent under the free space medium since the latter resembles the propagative medium, at least outside the scattering area. The associated additive component is the incident field which is contained in the total field detected in the ROI. This effect is very noticeable for the realistic finite dimensions of the scattering region considered in the aforementioned example. It is possible, of course, to remove this vulnerability via the adoption of large scattering domains or more complicated materials; but these findings clearly showcase an advantage of the differential sensing approach and pave the way for implementations based on quite simple media, e.g., single layer materials or metasurfaces. Indeed, in additional computer simulations (results not shown) involving simple media such as a single layer of scatterers, we consistently observed the aforementioned vulnerability to ``free space'' attacks which was not, however, present at all in the new differential methods. On the other hand, the single medium reconstruction in Fig.~\ref{fig_5} is clearly superior, relative to the one in the differential method (Fig.~\ref{fig_4}). This is expected since in the single medium framework the realizable waveforms are due to both the holographic source and the induced source in the scattering medium. In contrast, in the differential method the scattered waveforms are only due to the induced source. It follows that the space of such realizable fields is larger in the single medium case than in the scattering-based, difference-based method. Accordingly, the corresponding SVD of the response matrices in the differential method have smaller effective signal spaces (corresponding to statistically significant singular values) relative to those of the single medium case. As a result, greater accuracy can be accommodated in the single medium case, since the latter has intrinsically more resolution. The challenge is, precisesly, that some of the information-carrying components are non-scatterer specific and are therefore susceptible to the aforementioned ``free space'' leaking. These results showcase a tradeoff between resolution and degree of security and thereby illustrate a fundamental principle in physical layer security, namely the successful adoption of channel subutilization mechanisms in order to achieve guaranteed security at the expense of reduced information rate or capacity (in the present context, imaging resolution).

\subsection{Digital Encoding}
As a second case study, we consider the use of complex scattering media for the encryption of a digital signal such as a barcode. In the validation application, the user presents (to the checkpoint authority) one of the media, or a passcode to materialize it. The verifying party, which has access to the complementary medium, generates (through the relevant imaging apparatus) a set of queries, in the form of specific probing hologram realizations governed by another, third party, through which a companion set of (output) digital signals (barcodes) becomes materialized in real time in the ROI. It is assumed that access to the queries and correct outputs or responses (lookup table) is controlled by this third party. The latter can function as a virtual or remote, upon-request comparator. In an alternative scheme, it can distribute at checkpoint (e.g., via the cloud) the pertinent queries (holograms) to the intended user and the correct outputs to the enforcing authority, thereby enabling decision directly by such authority in real time. In this approach, statistically robust retrieval of the correct responses is indicative of a valid user. In contrast, poor statistical performance invalidates the user. The signal retrieval application has the same elements but in this case the output is the resulting barcode sequence itself.

 For consistency, we use in the following the same random media as before. As in the prior developments, we compare differential versus non-differential encryption methods.
Figure~\ref{fig_6}(a) shows the results for the single medium method, where the desired barcode is in gray. Different colors represent different cases: green for the correct medium (B), red for the eavesdropping medium (E), and black for free space. The bit error rate (BER), which is the number of incorrectly reconstructed bits divided by the total number of signal bits, is used to analyze these results. 

In digitizing the received analog image, if the absolute value of the reconstructed signal in a pixel is greater than 0.4, it is considered ``1''; otherwise, it is ``0''. This threshold of 0.4 was chosen for its good performance (via trial and error). In practice the threshold can be embedded into the recovery system. For instance, it is part of the query sent by the remote high-level third party. 
In the case considered in Fig.~\ref{fig_6}, the reconstruction is perfect when the correct scattering key (medium B) is selected. In the interception scenario in which eavesdropper medium E is used the error rate is equal to $53\%$ (reconstruction is almost completely random). Fig.~\ref{fig_6}(a) further showcases previously discussed vulnerabilities of the single-medium approach to free space reconstruction attempts. In the free space scenario, the reconstructed signal has considerable similarity to the original barcode, with an error rate of only $10.5\%$. Thus, at least for simple multiple scattering systems such as the ones considered in this study, a significant amount of information remains interceptable within the single-medium approach. 
As we illustrate next, the differential approach is quite effective in overcoming this deficiency. Figure~\ref{fig_6}(b) shows the corresponding reconstructions in the differential sensing approach. In this simulation the digital signal is encoded in the magnitude but the phase could have been adopted instead. In these plots, green, red, and black colors are associated with experiments using the correct background (B) and total media (T), the background medium (B) and eavesdropping medium (E), and the background medium (B) and free space as the second key, respectively. 
In this case, recovery with the correct media is quite accurate, with an error rate of only $2\%$. On the other hand, if only one of the correct media is adopted the reconstruction performance drops significantly. This corresponds to the unlikely scenario in which one of the required media is intercepted by an unintended user. In this experiment, the error rate under free space reconstruction is $47.5\%$, rendering the sought-after recovery almost totally random. If medium E is used the error rate is also high ($41.5\%$). 

\begin{figure}[H]
\centering
\includegraphics[width=15 cm]{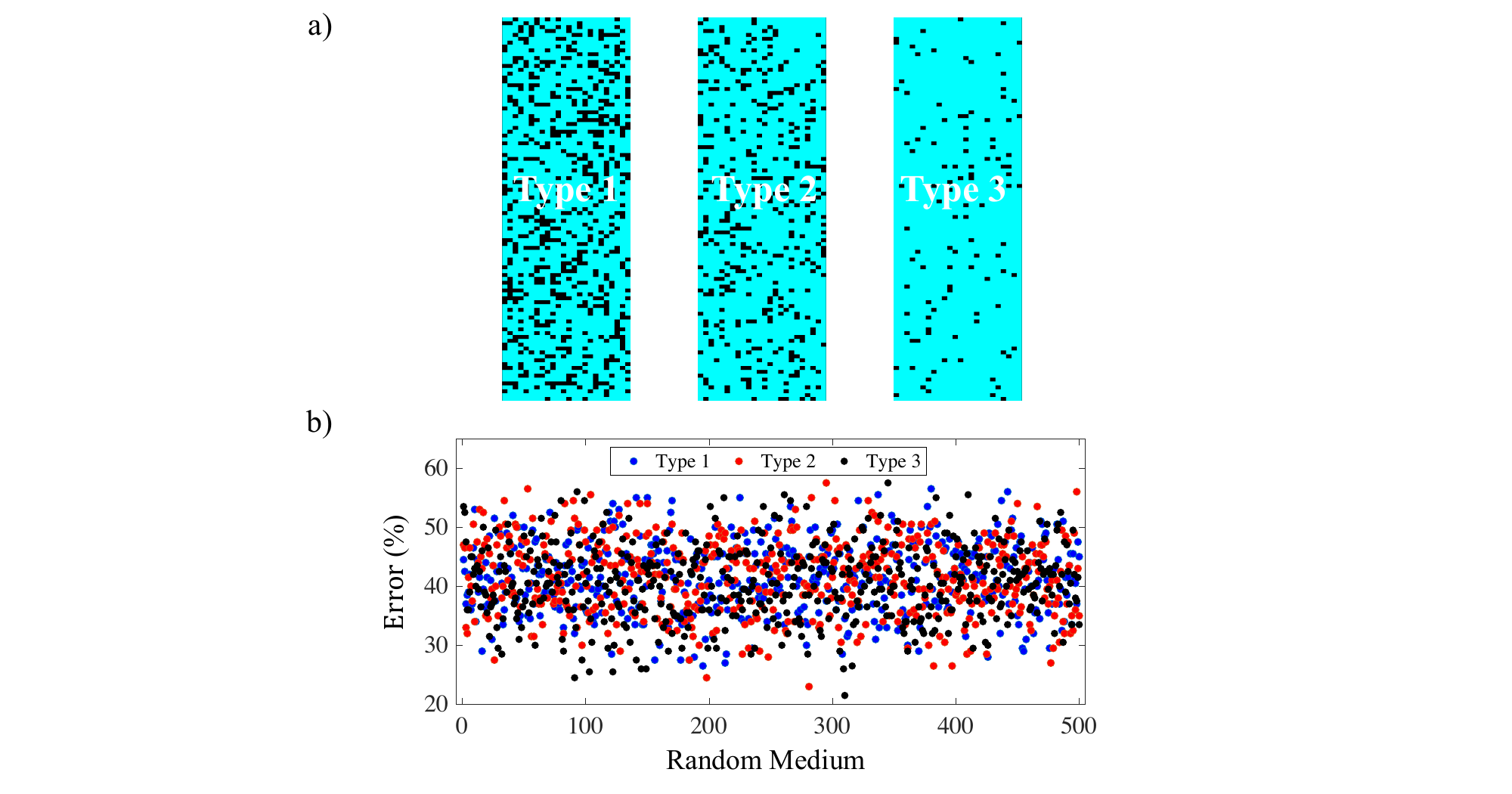}
\caption{ (a) Three different types of random media with varying densities of scatterers. (b) BER for 500 realizations of each medium type. }
\label{fig_6_stats}
\end{figure}

To further characterize these findings, we ran exhaustive simulations to determine the average BER of different eavesdropper media realizations. We considered, in particular, random realizations of media having different population probabilities to evaluate the impact of scatterer density on reconstruction performance. Figure~\ref{fig_6_stats}(a) illustrates typical realizations of eavesdropping media types ``1'', ``2'', and ``3'', corresponding to population probabilities of $5$, $15$, and $25\%$, respectively. Five hundred samples of each medium type were selected as the interceptor medium and the associated reconstruction was calculated for each case. The respective BERs are shown in Fig.~\ref{fig_6_stats}(b). The computed error rates are always larger than those calculated for the correct medium, as desired. An average BER of $40\%$ is obtained and, interestingly, the statistical characteristics are not affected by the scatterer density. Moreover, the smallest achieved BERs are in the vicinity of $20\%$, which is still distant (statistically distinguishable) from the correct medium rate of only $2\%$. 
In this interception analysis, it is assumed that the intruder has access to the correct background medium (B) (a pessimistic scenario, per se) and uses random keys as the total medium.

It is important to highlight that the observed insensitivity to the scatterer density is beneficial for the sought-after encryption. In particular, this result implies that the intruder cannot rely on key guessing strategies based on the adoption of a particular type of medium, namely, sparse media, dense material, and so, since the randomness associated to different kinds of materials is statistically the same. After understanding the effect of population of scatterers on the performance of the brute-force experiments, we turn our focus to similarity studies. 

\begin{figure}[H]
\centering
\includegraphics[width=15 cm]{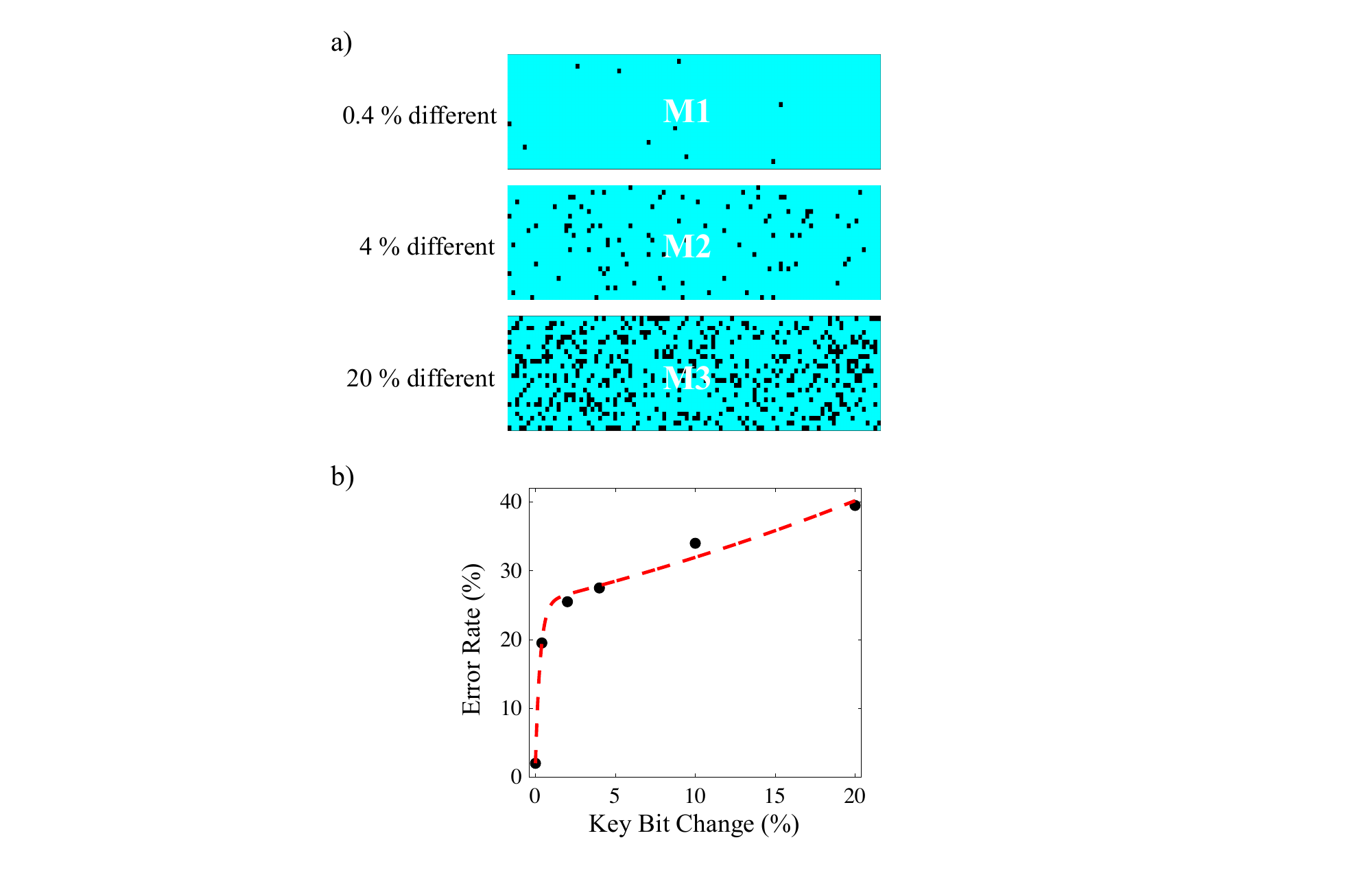}
\caption{ Interception analysis for media with different degrees of similarity relative to the correct medium. (a) Typical error bits for media M1, M2, and M3 having $0.4\%$, $4\%$, and $20\%$ difference percentage, respectively. (b) Error rate as a function of difference percentage.}
\label{fig_7a}
\end{figure}

Another important way to characterize the robustness of the proposed system is to test various random media having different degrees of similarity with respect to the correct medium. Figure ~\ref{fig_7a}(a) illustrates representative media with difference ratios of $0.4\%, \; 4\%, \; 20\%$ (corresponding to $99.6\%, \; 96\%, \; 80\%$ similarity percentage, respectively). The figure shows the error bits in these media, which we refer to as M1, M2, and M3, respectively. 
 Here it is important to emphasize that for the media considered in this section, involving a large number of scattering centers, the probability that an eavesdropper employs keys with high similarity ratios is (astronomically) negligible. Figure \ref{fig_7a}(b) illustrates the statistical performance of typical realizations of these media in the form of a plot of reconstruction error versus difference percentage. In these simulations we adopted the same barcode as in Fig.~\ref{fig_6}. Clearly the error rate increases in a steep manner as the material difference varies slightly from zero, and levels off for larger difference values as it approaches the asymptotic limit of 50\% which is indicative of totally random guessing of the barcode.

\subsection{Wavefield Nulling-Based Approach} 
We conclude this section with a discussion of an alternative approach that is relevant for both validation and secure communication. In this methodology, the hologram is designed such that the generated field at the sensing aperture vanishes only when the correct micromedia are used in the recovery phase. This nulling effect can serve as a validation mechanism since it occurs only with the correct media. This method may be implemented in a cost-effective manner via a bucket detector and is relevant for both the non-differential-sensing and the differential-sensing encryption methods. In either case, the idea is to employ the null-subspace ``source'' singular functions associated to the singular system of the relevant scattering response (for the randomly chosen media) as a basis for the construction of the pertinent holographic source. 

In the differential sensing case, the (null-subspace) singular functions $\rho_n^{(s)}$, for $\sigma_n^{(s)}<\varepsilon$ where $\varepsilon$ is a small tolerance parameter, are selected, and the holographic source adopted in the hologram synthesis ($\rho_h$) is expressed as  a superposition of said functions, e.g., 
\begin{equation}
    \rho_h = \sum_{\sigma_n^{(s)} < \varepsilon} C_n \rho_n^{(s)} .\label{aug_15_2024_1}\end{equation}
   The 2 norm or signal energy of this source is 
   \begin{equation}
       E_{0}=\sum_{\sigma_n^{(s)}<\varepsilon} |C_n|^2 .\label{aug_16_2024_1}\end{equation}
   The corresponding scattered field 
   \begin{equation}
       \psi_s = \sum_{\sigma_n^{(s)}<\varepsilon} \sigma_n^{(s)} C_n \psi_n^{(s)} \label{aug_15_2024_2}\end{equation}
       and consequently for small $\varepsilon \simeq 0$ the signal energy $E$ of the scattered field as measured at the sensing aperture, 
\begin{equation}
    E = \sum_{\sigma_n^{(s)} < \varepsilon} \left(\sigma_n^{(s)}\right)^2 |C_n|^2 < \varepsilon^2 E_{0}, \label{aug_15_2024_3}
\end{equation}

       is small, as desired.  
   
\begin{figure}[H]
\centering
\includegraphics[width=15 cm]{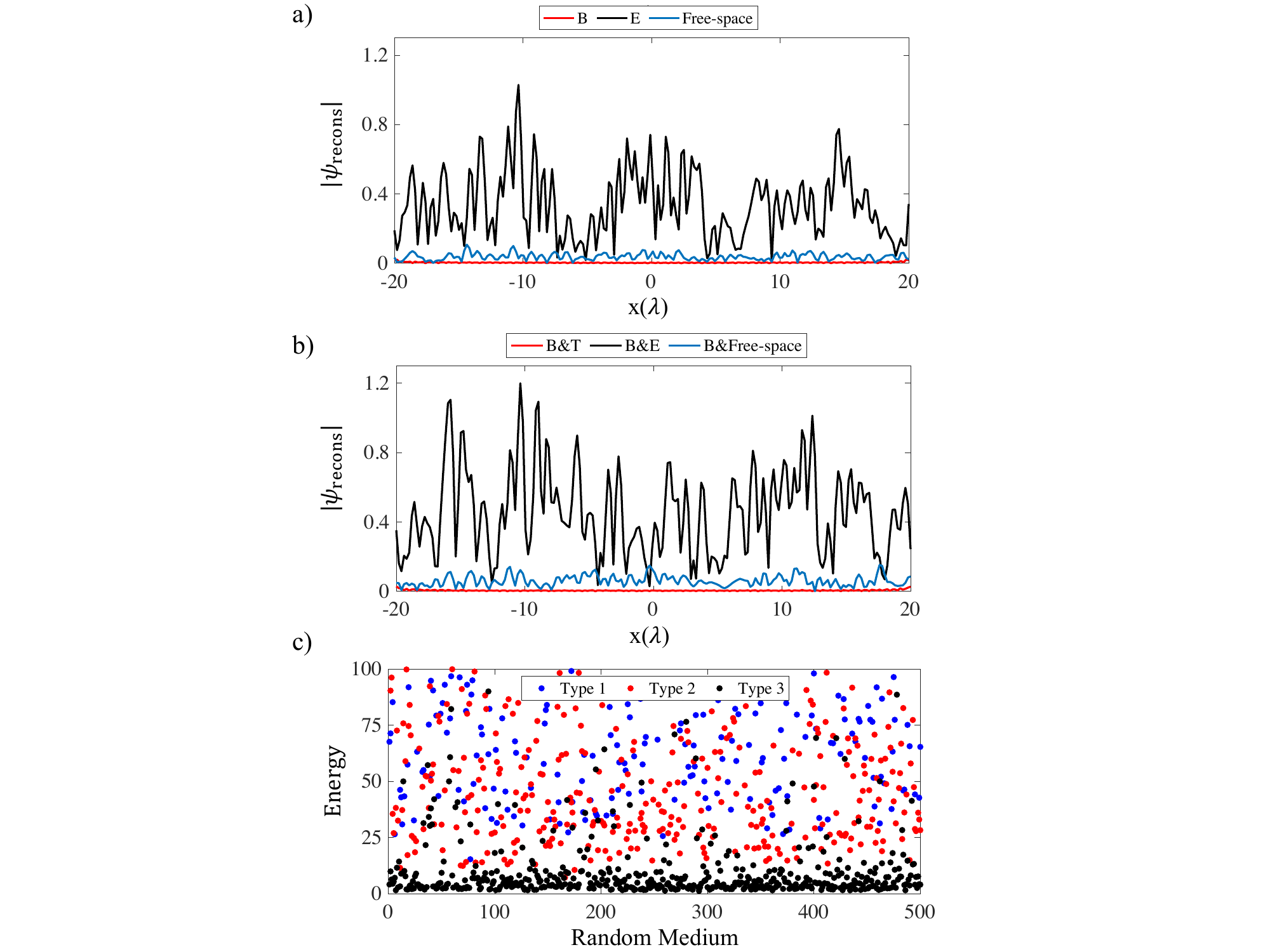}
\caption{(a), (b) Amplitude of the nulling-approach reconstructed field for the single-medium and differential methods. (c) Signal energy or intensity for 500 random samples corresponding to media types 1, 2, and 3. The results are plotted versus sample index. }
\label{fig_7}
\end{figure} 

\begin{figure}[H]
\centering
\includegraphics[width=15 cm]{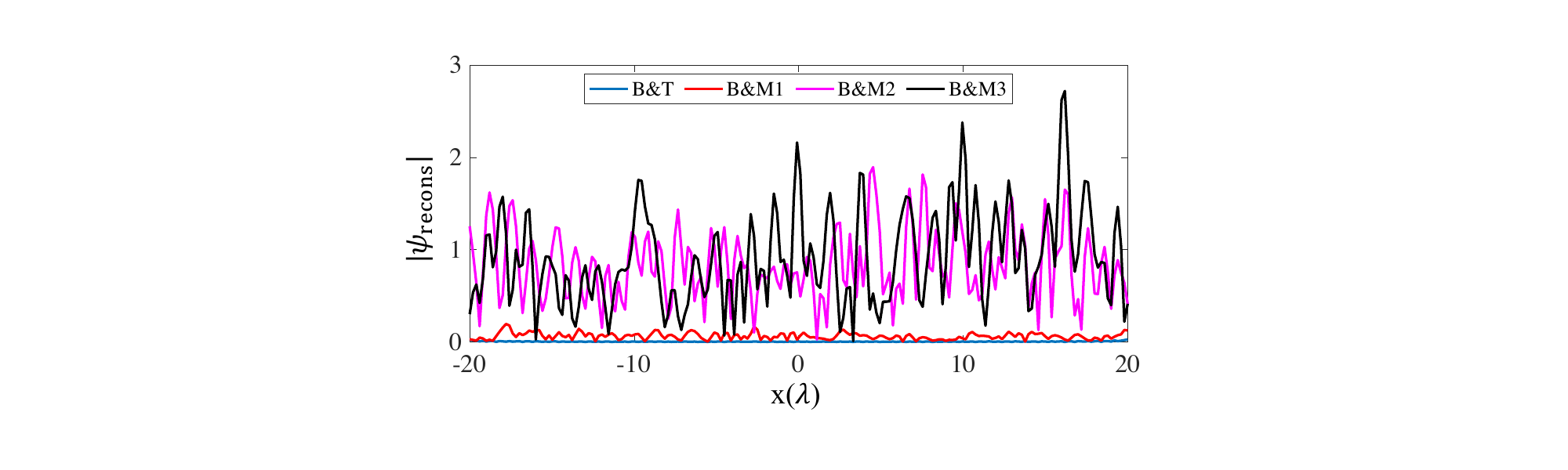}
\caption{Amplitude of the reconstructed signal for typical samples of intrusion media M1, M2, and M3. The amplitude for the correct medium T is very small, so it appears as a horizontal line coinciding with the $x$ axis.}
\label{fig_8}
\end{figure}

Figure \ref{fig_7}(a) shows the results of a representative single-medium example. As shown in the figure, the amplitude of the reconstruted waveform exhibits the expected nulling for the correct medium B (red line). The figure also shows a typical reconstruction (black line) based on eavesdropping medium (E) under the assumption that the eavesdropper has access to hologram $h$. No nulling occurs for such eavesdropping medium, as desired. Importantly, the change in amplitude is quite drastic. We found it to be between 4 and 5 orders of magnitude larger in the eavesdropping medium (E) than in the correct medium simulation. Also shown (blue line) is the corresponding plot if the eavesdropper assumes free space as the correct medium. No nulling occurs in that case either, as expected. Figure \ref{fig_7}(b) provides representative plots for the differential sensing approach. The results assume the extreme situation in which the eavesdropper has access to both hologram $h$ and the backbround medium B. As desired, the sought-after wavefield nulling is achieved for the correct total medium T, but no nulling occurs either under eavesdropping medium (E) or free space acting as the second medium, as desired. 
To examine the resilience to brute force attacks, samples of the three types of random scattering media considered earlier (types 1, 2, and 3) were generated and the energy or intensity of the respective reconstructed signals were computed. Figure \ref{fig_7}(c) shows the corresponding results, which clearly show that the intensities tend to be larger for dense media. For sparse media the intensities tend to be smaller but they are still orders of magnitude above the small value achieved under the correct total medium T for which the sought-after nulling effect occurs. Figure \ref{fig_8} shows plots of the reconstructed signal amplitude for a number of typical samples of these eavesdropping media. These results are encouraging since for media consisting of a large number of scatterers the dense media types are overwhelmingly dominant, statistically. Therefore both the randomly generated encryption media T and the vast majority of the eavesdropping media that would come up randomly in brute force attacks is expected to fall in this category. Clearly the sought-after contrast in signal energy or intensity tends to be very large between the correct medium T and the incorrect eavesdropping media within this dominant category, and adoption by the eavesdropper of the correct key (T) is unlikely as long as the search space is astronomically large as is the case for media consisting of a large number of scattering centers. We conclude by mentioning that, within a dynamic setting in which triplets of hologram $h$ and media B and T vary, then sequences of measurements revealing nulling or no-nulling can be adopted for the encoding of digital signals, or alternatively blocks of media for which nulling or no-nulling occurs can be selectively drafted for the storage or communication of secret keys and other purposes. Thus the derived nulling-based approach offers many interesting possibilities and within this approach both single-medium and differential methods exhibit good performance characteristics. 

\section{Conclusions}

In this paper we developed a new approach for the holographic encryption of signals that is based on the use of two complex and random scattering media. The derived methodology, termed ``differential sensing'', is based on the encoding of wave information in the scattered field due to a complex scatterer that is embedded in a complex background. The proposed approach is coherent and thus relies on holography for both the launching of the desired probing fields as well as for the sensing itself, for which a number of holographic measurement techniques were proposed. Within this approach the presence of both required media, which we called ``background'' (B) and ``total'' (T) media, is necessary for the formation of the relevant decryption imaging channel, through which the sought-after information-carrying plaintext image can be recovered from the given ciphertext hologram ($h$). In this methodology the cryptographic key is formed by two complex and random media and is thus harder to find than in the conventional single-medium approach, thereby rendering enhanced security. The derived approach employs a wavefield-inversion approach for the encryption step giving rise to the generation of (ciphertext,key) pairs and, importantly, this step can be based on either real or simulated data. Thus in contrast to prior work in this area, the computational burden is placed on the encryption step which may be carried out by a specialized party or facility while the corresponding decryption is done in real time, analogically, via physical imaging in the presence of the correct decryption media. This context is motivated by applications requiring real-time validation of a customer  in front of a suitable authority, but the concept is also relevant for conventional data storage and communications where it speeds up signal retrieval and frees the intended receiver from unnecessary computational burden. The proposed differential sensing method was illustrated with computer simulations corresponding to three different applications: image encryption of analog signals, image encryption of digital signals, and nulling-based methods for validation and communication where, depending on the application, the required complex media may be adopted either as key or as information carrier. The obtained results clearly illustrate the potential practical applicability of the methods developed in the paper and showcase the advantages of the differential sensing method relative to the conventional single-medium approach. 

\section{Acknowledgments}
The authors wish to thank Northeastern University’s Research Computing team for providing resources for high-performance computing through Discovery Cluster.
\bibliography{main}
 \bibliographystyle{ieeetr}
\end{document}